\documentclass{llncs}
\usepackage[a4paper,hmargin=1.6in,vmargin=1.25in]{geometry}
\usepackage[sectionbib,square, numbers, comma, sort&compress]{natbib}  
\usepackage[T1]{fontenc}
\usepackage{comment}
\usepackage{amsmath, environ}
\usepackage{paralist}
\usepackage{amssymb}
\usepackage{mathrsfs}
\usepackage{amsfonts}
\usepackage{stackrel}
\usepackage{algorithm2e}
\usepackage{enumitem}
\usepackage{hyperref}

\hypersetup{urlcolor=blue, colorlinks=true}  
\newcounter{namedefctr}
\newenvironment{namedef}[1]{
   \par
   \noindent        
   \refstepcounter{namedefctr}
   \textbf{Definition \thenamedefctr \enskip {\em #1:}}
   \itshape \selectfont
   \noindent
   }{\par}  
\numberwithin{namedefctr}{chapter}
\newcommand{\pr}{\mathbf{Pr}}
\newcommand{\expc}{\mathbf{E}}

\newcommand{\pf}{\noindent{\em Proof. }}
\newcommand\numberthis{\addtocounter{equation}{1}\tag{\theequation}}
\newcommand{\pfc}{\noindent{\em Proof of correctness. }}
\newcommand{\cn}{\noindent {\em Construction.}}

\allowdisplaybreaks
\begin{document}
\setlength{\abovedisplayskip}{0pt}
\setlength{\belowdisplayskip}{0pt}
\setlength{\abovedisplayshortskip}{0pt}
\setlength{\belowdisplayshortskip}{0pt}
\setlength{\jot}{-1pt}
\setlist[itemize]{noitemsep, topsep=0pt}
\setlist[enumerate]{noitemsep, topsep=0pt}
\setlist{nolistsep}

\setlist[description]{%
noitemsep,
topsep=0pt,              
  font=\normalfont\itshape 
}

\pagestyle{headings}

\mainmatter

\title{Derandomized Construction of Combinatorial Batch Codes}


\author{Srimanta Bhattacharya}

\institute{Centre of Excellence in Cryptology,\\
Indian Statistical Institute, Kolkata,\\
India 700108\\
\email{mail.srimanta@gmail.com}\\
}

\maketitle

\begin{abstract}
{\em Combinatorial Batch Codes} (CBCs), replication-based variant of {\em Batch Codes} introduced by Ishai {\em et al.} in \cite{YuKuOsSa}, abstracts the following data distribution problem: $n$ data items are to be replicated among $m$ servers in such a way that any $k$ of the $n$ data items can be retrieved by reading at most one item from each server with the total amount of storage over $m$ servers restricted to $N$. 
Given parameters $m, c,$ and $k$, where $c$ and $k$ are constants, one of the challenging problems is to construct $c$-uniform CBCs (CBCs where each data item is replicated among exactly $c$ servers) which maximizes the value of $n$. In this work, we present explicit construction of $c$-uniform CBCs with $\Omega(m^{c-1+{1 \over k}})$ data items. The construction has the property that the servers are almost {\em regular}, i.e., number of data items stored in each server is in the range $\left[{nc \over m}-\sqrt{{n\over 2}\ln (4m)}, {nc \over m}+\sqrt{{n \over 2}\ln (4m)}\right]$. The construction is obtained through better analysis and derandomization of the randomized construction presented in \cite{YuKuOsSa}. Analysis reveals almost regularity of the servers, an aspect that so far has not been addressed in the literature. The derandomization leads to explicit construction for a wide range of values of $c$ (for given $m$ and $k$) where no other explicit construction with similar parameters, i.e., with $n = \Omega(m^{c-1+{1 \over k}})$, is known. Finally, we discuss possibility of parallel derandomization of the construction.
\end{abstract}

\section{Introduction}
\subsection{Background}
\noindent{\bf{\em Batch codes.}} An $(n, N, k, m)$-{\em batch code} (or $(n, N, k, m, t=1)$-{\em batch code}\footnote{In \cite{YuKuOsSa}, batch codes were defined for general $t$. However, in this work we solely consider $t=1$ case as that seems to capture crux of the problem.}) abstracts the following data distribution problem: $n$ data items are to be distributed among $m$ servers in such a way that any $k$ of the $n$ items can be retrieved by reading at most one item from each server and total amount of storage required for this distribution is bounded by $N$. 
Batch codes were introduced in \cite{YuKuOsSa}, and their primary motivation was to amortize computational work done by the servers during execution of {\em private information retrieval} protocol by batching several queries together while limiting total storage (see \cite{YuKuOsSa} for more details). It is also easy to see from the above description that these codes can have potential application in a distributed database scenario where the goal is to distribute load among the participating servers while optimizing total storage.\par
On the theoretical side batch codes closely resemble several combinatorial objects like {\em expanders}, {\em locally decodable codes}, etc.,  and there is also similarity with Rabin's {\em information dispersal}. However, there are fundamental differences of batch codes with these objects, especially as far as setting of parameter values are concerned, and it seems difficult to set up satisfactory correspondences with these objects. This dichotomy makes batch codes unique and very interesting objects.\par
 
\noindent{\bf{\em Combinatorial batch codes (CBCs).}} These are replication based batch codes; each of the $N$ stored data items is a copy of one of the $n$ input data items. So, for CBCs, encoding is assignment (storage) of items to servers and decoding is retrieval (reading) of items from servers. This requirement makes CBCs purely combinatorial objects. As combinatorial objects CBCs are quite interesting, they have received considerable attention in recent literature(\cite{PaStWe, BrKiMeSc, BuTu2, BuTu3, BuTu5, GaSi, BhRuRo, BaBh}). Before proceeding further, we introduce the formal framework for our study of CBCs.\par

We represent an $(n, N, k, m)$-CBC $\mathcal{C}$ as a bipartite graph $\mathcal{G}_{\mathcal{C}} = (\mathcal{L}, \mathcal{R}, \mathcal{E})$. Set of left vertices $\mathcal{L}$ represents $\lvert \mathcal{L}\rvert = n$ data items with vertex $u_i \in \mathcal{L}$ representing data item $x_i, 1 \leq i \leq n$, and set of right vertices $\mathcal{R}$ represents $\lvert \mathcal{R} \rvert = m$ servers with vertex $v_j \in \mathcal{R}$ representing server $s_j, 1\leq j \leq m$. Hence, $(u_i, v_j) \in \mathcal{E}$ is an edge in $\mathcal{G}_{\mathcal{C}}$ if the data item $x_i$ is stored in server $s_j$. Since the total storage is $N$ so it follows that $\sum_{u \in \mathcal{L}}deg(u)= \sum_{v \in \mathcal{R}}deg(v)  = \lvert \mathcal{E} \rvert = N$, where $deg(.)$ is the degree of a vertex in $\mathcal{G}_\mathcal{C}$. Now, it can be observed that any subset $\{x_{i_1}, \ldots, x_{i_k}\}$ of $k$ distinct data items can be retrieved by reading one item each from $k$ distinct servers iff the sets $\Gamma(u_{i_1}), \ldots, \Gamma(u_{i_k})$ has a {\em system of distinct representatives} (SDR), where $\Gamma(u_r), r \in\{1, \ldots, n\}$ is the {\em neighbourhood} of the vertex $u_r \in \mathcal{L}$, i.e., $\Gamma(u_r) = \{v \in \mathcal{R}, (u_r, v) \in \mathcal{E}\}$\footnote{In the sequel, we will require extension of the definition of neighbourhood of a vertex to neighbourhood of a subset. More formally, given $S \subset \mathcal{L}$, we denote by $\Gamma(S)$ the set $\{v \in \mathcal{R} \vert \exists u \in S, (u, v) \in \mathcal{E}\}$}. According to {\em Hall's theorem} for SDR (cf. \cite{Bol3}, pp. 6) this is equivalent to the condition that union of any $j$ sets $\Gamma(u_{i_1}), \ldots \Gamma(u_{i_j})$, $\{u_{i_1} \ldots u_{i_j}\} \subset \mathcal{L}$ contains at least $j$ elements for $1 \leq j \leq k$. These considerations lead naturally to the following theorem of \cite{PaStWe} which can also be thought as defintion of a CBC.

\begin{theorem}[\cite{PaStWe}]
A bipartite graph $\mathcal{G}_{\mathcal{C}}=(\mathcal{L}, \mathcal{R}, \mathcal{E})$ represents an $(n, N, k,\\ m)$-CBC if $\lvert \mathcal{L} \rvert = n$, $\lvert \mathcal{R} \rvert = m, \lvert \mathcal{E} \rvert = N$ and union of every collection of $j$ sets\\ $\Gamma(u_{i_1}), \ldots, \Gamma(u_{i_j})$, $\{u_{i_1}, \ldots, u_{i_j}\} \subset \mathcal{L}$ contains at least $j$ elements for $1 \leq j \leq k$.
\label{thm2}
\end{theorem}
From now on, we will identify the graph $\mathcal{G}_{\mathcal{C}}=(\mathcal{L}, \mathcal{R}, \mathcal{E})$ with an $(n, N, k, m)$-CBC, and omit the subscript $\mathcal{C}$ as it will not cause any trouble. A CBC $\mathcal{G}= (\mathcal{L}, \mathcal{R}, \mathcal{E})$ is called {\em $c$-uniform} if for each $u \in \mathcal{L}$, $deg(u) = c$, and it is called {\em $l$-regular} if for each $v \in R, deg(v) = l$\footnote{Here the terminology is in keeping with the representation of a CBC as a {\em set-system} in some of the previous works, where the set $\mathcal{R}$ is treated as ground set and the multi-set $\{\Gamma(u_1) \ldots \Gamma(u_n)\}, u_i \in \mathcal{L}$ is the collection of subsets.}. Two optimization problems related to CBCs have been addressed in the literature:\begin{inparaenum}[(i)]\item given $n, m, k$  find minimum value of $N$ attained by a CBC (not necessarily uniform or regular) , and provide explicit construction of corresponding extremal CBCs; \item given $m, c, k$, find maximum value of $n$, denoted as $n(m, c, k)$, attained by a $c$-uniform CBC (not necessarily regular), and provide explicit construction of corresponding extremal CBCs.  \end{inparaenum} In this work we will consider the latter problem in setting of parameters where $c$ and $k$ are constants while $m$ is variable.\par
At this point, it may be observed that though the definition and the considered problem draws some similarity with those of bipartite expanders, especially the unbalanced expanders \cite{GUV09}, there are important differences between these two cases as well. On the similarity side,  both are bipartite graphs with constant left-degree; in both the cases, it is required that every subset of vertices $\mathcal{L}$ of up to a specified size should have neighbourhood of specified sizes, and it is desirable that $\lvert \mathcal{L}\rvert >> \lvert \mathcal{R}\rvert$. However, in the case of unbalanced expanders, the goal is to stretch the expansion of subsets\footnote{For a set $S$, expansion of $S$ is $\Gamma(S) \over S$.} (of specified sizes) of $\mathcal{L}$ as close to the left-degree as possible. Whereas, in case of CBCs, expansion of $1$ is sufficient and it is more important to make $\lvert \mathcal{L}\rvert$ as large as possible with respect to $\lvert \mathcal{R}\rvert$. Also important is the fact that the parameter $k$ is a constant in case of CBCs (within our setting of parameters), whereas for expanders $k$ varies with $n$. These differences make the (desirable) parameters in these two cases essentially unrelated. So, it seems unlikely that the existing constructions of unbalanced expanders can be immediately used for construction of CBCs where $c$ and $k$ are constants.\par
Before discussing existing results and our contribution we briefly mention the notion of `explicit' construction of a combinatorial object in general and CBCs in particular as that will be crucial to our discussion and result.\par
\noindent{\bf{\em Explicit construction.}} Construction of a combinatorial object with desirable properties is computation of a representation of the object and is tied with the resources used for the computation. In the literature, those constructions which require practically feasible amount of resources, such as polynomial time or logarithmic space are termed explicit. This can be contrasted with exhaustive search of a combinatorial object whose existence has been proven (e.g. by probabilistic argument); the search is done in the space of the object and requires infeasible amount of resources (e.g. exponential time). The notion of explicitness we will adhere to in this work is polynomial time constructibility, which requires that the time required for the construction be bounded by a polynomial in the size of the representation. Explicitness is further classified as following\footnote{Though the classification is with respect to polynomial time constructibility, it is applicable for other feasible resource bounds as well.}.
\begin{description}
\item [Globally explicit.] In this case the whole object is constructed in time polynomial in the size of the object. For example, a globally explicit construction of CBC $(\mathcal{L}, \mathcal{R}, \mathcal{E})$ would list all the members of $\mathcal{E}$ in $poly(\lvert \mathcal{E}\rvert)$ time.
\item [Locally explicit.] In this case the idea is to have quick local access to the object. More formally, for a desirable combinatorial object $G$, locally explicit construction of $G$ is an algorithm which given an index of size $\log (\lvert G \rvert)$, outputs the member of $G$ with the given index (or does some local computation on the member) in time $polylog(\lvert G \rvert)$. This is more specialized notion and depends on the context. For example, a locally explicit construction of a $c$-uniform CBC $(\mathcal{L}, \mathcal{R}, \mathcal{E})$ would list the neighbourhood of a vertex $v \in \mathcal{L}$ in time $poly(\log(\lvert \mathcal{L} \rvert), c)$ given the index of $v$ (which is of size $\lvert \mathcal{L} \rvert$). It can be seen that locally explicit construction is a stronger notion than globally explicit construction and is always desirable as it is useful for algorithmic applications. In fact, common notion of construction of combinatorial objects (e.g. using various algebraic structures) falls in this category. 
\end{description}
Now, we state relevant existing results for CBCs and subsequently discuss our contribution.
\subsubsection{Existing Results}
\begin{itemize}
\item In \cite{YuKuOsSa}, the authors have shown, {\em inter alia}, that $n(m, c, k) = \Omega(m^{c-1})$; this bound was obtained using probabilistic method.
\item In \cite{PaStWe}, the authors have refined the above estimate using the {\em method of deletion} (another probabilistic technique, see \cite{AS}) to $n(m, c, k) = \Omega(m^{{ck \over k-1}-1})$. They have also shown through explicit construction that $n(m,k-1,k) = (k-1)\binom{m}{k-1}$, and $n(m, k-2, k) = \binom{m}{k-2}$.  
\item In \cite{BaBh}, it was shown that $n(m, c, k) = O(m^{c-{1 \over 2^{c-1}}})$ for $7 \leq k$, and $3 \leq c \leq k-\lceil\log k \rceil-1$, and for $k-\lceil \log k \rceil \leq c \leq k-1$, it was shown through explicit construction that $n(m, c, k) = \Theta(m^c)$. For $c=2$ case, the lower bound of (ii) was improved (through explicit construction) to $n(m, 2, k)= \Omega(m^{\frac{k+1}{k-1}})$ for all $k \geq 8$ and infinitely many values of $n$.
\item In \cite{BuTu5}, the authors improved the general upper bound to show that $n(m, c, k) = O(m^{c-1+{1 \over \lfloor {k \over c+1}\rfloor}})$ for $c \leq {k \over 2} -1$.
\end{itemize}
All the constructions mentioned above are locally explicit.
\subsubsection{Our Contribution}
We address the question of explicit construction of uniform and regular CBCs. Here it is noteworthy that the aspects of regularity and uniformity have not been addressed together so far in the literature; addressing both the properties together is interesting and important from theoretical as well as practical point of view. We provide a globally explicit construction which is uniform and almost regular. In particular, our result is the following.
\begin{theorem}
\label{cbcbound1}
Let $c, k$ be positive constants. Then for all sufficiently large $m$, there exists $c$-uniform $(n, cn, k, m)$-CBC, where $n = \Omega(m^{c-1+{1 \over k}})$, and number of items in each server is in the range 
$\left[{nc \over m}-\sqrt{{n \over 2}\ln (4m)}, {nc \over m}+\sqrt{{n \over 2}\ln (4m)}\right]$. Moreover, the CBC can be globally constructed in $poly(m)$ time.
\end{theorem}
We use the construction of \cite{YuKuOsSa} and derandomize it using the {\em method of conditional expectation} (see \cite{AS}), and analyze it in greater detail. The analysis shows almost regularity of the construction and also improves the exponent of the lower bound ($\Omega(m^{c-1+{1 \over k}})$ as opposed to $\Omega(m^{c-1})$ of \cite{YuKuOsSa}). Though the improved exponent is inferior to the one obtained in \cite{PaStWe} ($\Omega(m^{c-1+{c \over k-1}})$), the importance of the construction lies in its almost regularity and {\em explicitness} which are not known for the construction of \cite{PaStWe}. In fact, apart from the range $k-\lceil \log k \rceil \leq c \leq k$, where $n = \Theta(m^c)$ is achieved (see \cite{PaStWe} for $c = k-1$ and $k-2$ and \cite{BaBh} for the remaining values), and for the $2$-uniform case (\cite{BaBh}) there is no explicit construction in the literature. So, our construction serves to fill the void to some extent.\par 

To describe our construction, we give an algorithm which, for given positive integers $k, c$, and sufficiently large $m$, runs in time $poly(m)$ and outputs the edges of a bipartite graph $(\mathcal{L}, \mathcal{R}, \mathcal{E})$, with $\lvert \mathcal{R} \rvert = m$, $\lvert \mathcal{L} \rvert= n =  {m^{c-1+{1 \over k}}\over 4k^{c+1}}$, satisfying the following conditions. 
\begin{enumerate}[label=(\alph*)]
\item Each left vertex in $\mathcal{L}$ has degree $c$ and each vertex in $\mathcal{R}$ has degree in the range $\left[{nc \over m}-\sqrt{{n \over 2}\ln (4m)}, {nc \over m}+\sqrt{{n \over 2}\ln (4m)}\right]$. 
\item Any subset of $i, 1\leq i\leq k,$ vertices in $\mathcal{L}$ has at least $i$ neighbours in $\mathcal{R}$.
\end{enumerate}
Note that there is a trivial non-explicit algorithm to construct the required graph which, given the input parameters, runs in time exponential in $m$; the algorithm searches the space of all possible bipartite graphs $(\mathcal{L}, \mathcal{R}, \mathcal{E})$, with $\lvert \mathcal{R} \rvert = m$, $\lvert \mathcal{L} \rvert= n =  {m^{c-1+{1 \over k}}\over 4k^{c+1}}$, and outputs one that satisfies the above two conditions.

In the proof of Theorem \ref{cbcbound1}, we will need the following version of Hoeffding's inequality.
\begin{theorem}({\bf Hoeffding's inequality}\cite{Hoe63})
\label{hoef}
Let $X_1, X_2, \ldots, X_n$ be independent random variables taking their values in the interval $[0, 1]$. Let $X = \sum_i X_i$. Then for every real number $a > 0$,$\pr\{\lvert X - \expc[X]\rvert \geq a\} \leq 2 e^{{-2a^2\over n}}$.
\end{theorem}
Also, given a set $\mathcal{S}$ and a positive integer $c (\leq \lvert \mathcal{S}\rvert)$, we will denote by $\binom{\mathcal{S}}{c}$, the set of all $c$ element subsets of $\mathcal{S}$.

\section{Proof of Theorem 2}
Proof of Theorem \ref{cbcbound1} is split into two parts. In the first part, we give probabilistic proof of existence (which essentially is also a randomized algorithm) of the CBC. In the second part, we derandomize the construction using the method of conditional expectation. This is a commonly used technique having its genesis in \cite{ES73} and was later on applied to prove many other derandomization results (e.g. \cite{Rag88, Spe94}). Informally, the method systematically performs a binary (or more commonly a $d$-ary) search on the sample space from where the randomized algorithm makes its choices and finally finds a good point.\par

\noindent{\bf{\em Proof of existence.}} We construct a bipartite graph $\mathcal{G}=(\mathcal{L}, \mathcal{R}, \mathcal{E})$, where $\mathcal{L}$ is the set $\{u_1, \ldots, u_n\}$ of $n$ left vertices, $\mathcal{R}$ is the set $\{v_1, \ldots, v_m\}$ of $m$ right vertices, and $\mathcal{E}$ is the set of edges, in the following manner. For each vertex in $\mathcal{L}$ we choose its $c$ distinct neighbours by picking randomly, uniformly, and independently a subset of $c$ vertices from $\mathcal{R}$, i.e., its neighbourhood is an independently and uniformly chosen element of $\binom{\mathcal{R}}{c}$. So, for $u \in \mathcal{L}, S' \subseteq \mathcal{R}$, 
\begin{align*}
\pr\left\{\Gamma(u) \subseteq S' \right\} = {\binom{\lvert S' \rvert}{c} \over \binom{m}{c}} \leq \left({\lvert S' \rvert \over m}\right)^c.
\end{align*}
Next, for a subset $S \subset \mathcal{L}, \lvert S \rvert =i, c+1\leq i \leq k$ and a subset $S' \subset \mathcal{R}, \lvert S' \rvert =i-1$, we say that event $Bad_{S, S'}$ has occured if $\Gamma(S) \subseteq S'$. So, we have
\begin{eqnarray}
\label{indep1}
\pr\{Bad_{S, S'}\}\leq \left({i-1 \over m}\right)^{ic},
\end{eqnarray}
using independence of the events $\Gamma(u) \subseteq S'$ for $u \in \mathcal{L}$. Now, our goal is to bound the probability of occurence of any $Bad_{S, S'}$, $S \subset \mathcal{L}$, $S' \subset \mathcal{R}$, and $c+1\leq i \leq k$. To this end, we have
\begin{align*}
\sum_{\substack{c+1\leq i\leq k}}\sum_{\substack{S \subset \mathcal{L}, \\ \lvert S\rvert =i}}~\sum_{\substack{S' \subset \mathcal{R}, \\ \lvert S'\rvert =i-1 }} \pr\{Bad_{S,S'}\} &\leq \sum \limits_{c+1\leq i\leq k}\binom{n}{i} \binom{m}{i-1} \left({i-1 \over m}\right)^{ic}\\
& \leq  \sum \limits_{1\leq i\leq k} n^i m^{i-1} \left({i-1 \over m}\right)^{ic}\\
& \leq  \sum \limits_{1\leq i\leq k} n^i m^{i-1} \left({k\over m}\right)^{ic}\\
& \leq  \sum \limits_{1\leq i\leq k} \left({1\over 4k}\right)^{i} ~~\text{since}~ n =  {m^{c-1+{1 \over k}}\over 4k^{c+1}}\\
& \leq  {1\over 4} \numberthis \label{expand1}
\end{align*} 
Next, for $u \in \mathcal{L}, v \in \mathcal{R}$ define the indicator random variable $X_{v}^{u}$ such that
$$
X_{v}^{u}=
\begin{cases}
1 & (u, v) \in \mathcal{E}\\
0 & \mbox{otherwise},
\end{cases} 
$$
and $X_v, v \in \mathcal{R}$, be a random variable denoting the degree of vertex $v$. Clearly $X_v = \sum \limits_{u \in \mathcal{L}} X_{v}^{u}$. Now, $\pr\{ X_{v}^{u} = 1\} = {c\over m}$. So, by linearity of expectation,
$$
\expc[X_{v}] = \expc\left[\sum\limits_{u \in \mathcal{L}}X_{v}^{u}\right]=\sum \limits_{u \in \mathcal{L}} \expc[X_{v}^{u}] = {nc\over m}.
$$
From the fact that neighbourhoods of vertices $u \in \mathcal{L}$ are chosen independently, it follows that the variables $X_{v}^{u}$ for  $u \in \mathcal{L}$, and a fixed $v \in \mathcal{R}$, are mutually independent. So, applying Theorem \ref{hoef} with $a = \sqrt{{n \over 2}\ln (4m)}$ we have
\begin{eqnarray}
\label{chreq}
\pr\left\{\lvert X_v-\expc[X_v]\rvert \geq \sqrt{{n \over 2}\ln (4m)} \right\} \leq 2(4m)^{-1}\leq {1 \over 2m}.
\end{eqnarray}
So, by union bound, probability that the event $\lvert X_v-\expc[X_v]\rvert \geq \sqrt{{n \over 2}\ln (4m)}$ occurs for some $ v, v \in \mathcal{R} $ is bounded by
\begin{eqnarray}
\label{balanced}
\sum \limits_{v\in \mathcal{R}}\pr\left\{\lvert X_v-\expc[X_v]\rvert \geq \sqrt{{n \over 2}\ln (4m)} \right\} \leq {1\over 2}.
\end{eqnarray}
Hence, from equations \eqref{expand1} and \eqref{balanced}, with probability at least $1 - ({1\over 4}+{1\over 2}) = {1 \over 4}$, none of the above events occur. \qed\par


\noindent{\bf{\em Derandomization.}} Before presenting the algorithm we derive expressions for expected number of $Bad_{S, S'}$ events and expected number of vertices $v \in \mathcal{R}$ for which $\lvert deg(v)-{nc \over m}\rvert >\sqrt{{n \over 2}\ln (4m)}$ conditional on fixed choices of $\Gamma(u_1), \ldots, \Gamma(u_t)$.  Then we show that if at $t$-th stage, $1 \leq t \leq n$, (having fixed $\Gamma(u_1), \ldots, \Gamma(u_{t-1})$) choice of $\Gamma(u_t)$ is made in such a way to minimize the sum of these two expectations then in the final graph, which is no longer random since all the neighbourhoods are fixed, there are no events $Bad_{S,S'}$ and no vertices $v \in \mathcal{R}$ for which $\lvert deg(v)-{nc \over m}\rvert >\sqrt{{n \over 2}\ln (4m)}$, i.e., no violations of conditions (a) and (b). So, the derandomization proceeds in $n$ stages; the beginning of stage $t$, neighbourhood of vertices $u_1, \ldots, u_{t-1} \in \mathcal{L}$ are fixed, and neighbourhood of $u_t$, $\Gamma(u_t) \in \binom{\mathcal{R}}{c}$ is fixed in such a way that minimizes the expected number of violations of conditions (a) and (b).  The algorithm (Algorithm \ref{alg1}) is immediate from these observations.\par  
First, we introduce indicator random variables $Y_{S, S'}$ corresponding to each event $Bad_{S, S'}$, i.e.,
$$
Y_{S, S'}=
\begin{cases}
1 & \mbox{if}~ \Gamma(S) \subseteq S'\\
0 & \mbox{otherwise}.
\end{cases} 
$$
Also, we define $Y = \sum_{\substack{c+1\leq i\leq k}}\sum_{\substack{S \subset \mathcal{L}, \\ \lvert S\rvert =i}}~\sum_{\substack{S' \subset \mathcal{R}, \\ \lvert S'\rvert =i-1 }}Y_{S,S'}$. By linearity of expectation we have
\begin{align*}
\expc[Y] &= \expc\left[\sum_{\substack{c+1\leq i\leq k}}\sum_{\substack{S \subset \mathcal{L}, \\ \lvert S\rvert =i}}~\sum_{\substack{S' \subset \mathcal{R}, \\ \lvert S'\rvert =i-1 }}Y_{S,S'}\right] \\
&= \sum_{\substack{c+1\leq i\leq k}}\sum_{\substack{S \subset \mathcal{L}, \\ \lvert S\rvert =i}}~\sum_{\substack{S' \subset \mathcal{R}, \\ \lvert S'\rvert =i-1 }}\expc[Y_{S,S'}]=\sum_{\substack{c+1\leq i\leq k}}\sum_{\substack{S \subset \mathcal{L}, \\ \lvert S\rvert =i}}~\sum_{\substack{S' \subset \mathcal{R}, \\ \lvert S'\rvert =i-1 }}\pr \{Y_{S,S'}\} \leq {1\over 4} ~\mbox{from \eqref{expand1}} 
\end{align*}
Let $\{u_1, u_2, \ldots, u_t\} \subseteq \mathcal{L}$, and $C_1, C_2, \ldots, C_t \in \binom{\mathcal{R}}{c} $ be fixed subsets such that $\Gamma(u_j) = C_j, 1\leq j\leq t$, and for the remaining vertices in $\mathcal{L}$, their neighbourhoods are chosen independently, and uniformly at random from $\binom{\mathcal{R}}{c}$. Let $S \subseteq \mathcal{L}, S' \subseteq \mathcal{R},\lvert S \rvert =i, \lvert S' \rvert = i-1$ be fixed subsets for some $i, c+1\leq i\leq k$, also let $W = S \cap \{u_1, u_2, \ldots, u_t\}$, $\lvert W \rvert = w$, and $\Gamma(W) = \emptyset$ for $W = \emptyset$. Then we have
\begin{align}
\label{speccondvio}
\nonumber
\expc[Y_{S, S'}\vert \Gamma(u_1)=C_1, \ldots, \Gamma(u_t) = C_t] &= \pr\{\Gamma(S) \subseteq S' \vert \Gamma (u_1)=C_1, \ldots, \ldots \Gamma(u_t) = C_t\}\\ 
&=
\begin{cases}
0 & \mbox{if}~ \Gamma(W) \nsubseteq S'\\
\left({\binom{i-1}{c}\over \binom{m}{c}}\right)^{i-w} & \mbox{otherwise}.
\end{cases}
\end{align}
So, by applying linearity of expectation and from above
\begin{flalign}
\label{condvio}
\nonumber
&\expc[Y\vert \Gamma(u_1)=C_1, \ldots, \Gamma(u_t) = C_t] \\
\nonumber
&=\sum_{\substack{c+1\leq i\leq k}}~\sum_{\substack{S \subset \mathcal{L}, \\ \lvert S\rvert =i}}~\sum_{\substack{S' \subset \mathcal{R}, \\ \lvert S'\rvert =i-1}}\expc[Y_{S,S'} \vert \Gamma (u_1)=C_1, \ldots, \ldots \Gamma(u_t) = C_t]\\
\nonumber
&=\sum_{\substack{c+1\leq i\leq k}}~\sum_{\substack{S \subset \mathcal{L}, \\ \lvert S\rvert =i}}~\sum_{\substack{S' \subset \mathcal{R}, \\ \lvert S'\rvert =i-1}}\pr\{\Gamma(S) \subseteq S' \vert \Gamma (u_1)=C_1, \ldots, \ldots \Gamma(u_t) = C_t\}\\
&=\sum_{\substack{c+1\leq i\leq k}}~\sum_{\substack{S \subset \mathcal{L}, \\ \lvert S\rvert =i}}~\sum_{\substack{S' \subset \mathcal{R}, \\ \lvert S'\rvert =i-1\\ \Gamma(W) \subseteq S' }}\left({\binom{i-1}{c}\over \binom{m}{c}}\right)^{i-w}.
\end{flalign}
Next, corresponding to each vertex $v \in \mathcal{R}$ we introduce an indicator random variable $Z_v$ such that
$$
Z_v=
\begin{cases}
1 & \lvert deg(v)-{nc \over m}\rvert >\sqrt{{n \over 2}\ln (4m)}\\
0 & \mbox{otherwise},
\end{cases} 
$$
and define $Z = \sum\limits_{v \in \mathcal{R}}Z_v$. So, by linearity of expectation we have 
\begin{align*}
\expc[Z] = \expc\left[\sum\limits_{v \in \mathcal{R}}Z_v\right] = \sum\limits_{v \in \mathcal{R}}\expc[Z_v]=\sum\limits_{v \in \mathcal{R}}\pr\{Z_v = 1\}\leq {1\over 2} ~\mbox{from \eqref{balanced}} 
\end{align*}
Like in the previous case, our goal is to estimate $\expc[Z\vert \Gamma(u_1)=C_1, \ldots, \Gamma(u_t) = C_t]$ by estimating $\expc[Z_v\vert \Gamma(u_1)=C_1, \ldots, \Gamma(u_t) = C_t]$ for each $v \in \mathcal{R}$. For a fixed $v \in \mathcal{R}$, let $l = \lvert \{u_i\vert v \in \Gamma(u_i), 1\leq i \leq t\}\rvert$.
Let $\alpha ={nc \over m} -\sqrt{{n \over 2}\ln (4m)} $, and $\beta ={nc \over m} +\sqrt{{n \over 2}\ln (4m)}$. Then we have
\begin{align}
\label{degvio}
\nonumber
&\expc[Z\vert \Gamma(u_1)=C_1, \ldots, \Gamma(u_t) = C_t] \\
\nonumber
&=\sum\limits_{v \in \mathcal{R}}\expc[Z_v\vert \Gamma(u_1)=C_1, \ldots, \Gamma(u_t) = C_t] \\
\nonumber
&= \sum\limits_{v \in \mathcal{R}}\pr\{deg(v) < \alpha -l ~~\mbox{or}~~ deg(v)> \beta-l \vert \Gamma(u_1)=C_1, \ldots, \Gamma(u_t) = C_t\}\\
\nonumber
&= \sum\limits_{v \in \mathcal{R}}\Bigl(\sum\limits_{i=0}^{i=\alpha-l-1} \binom{n-t}{i}\left({c \over m}\right)^i\left(1-{c \over m}\right)^{n-t-i}+\sum\limits_{i=\beta-l+1}^{n-t} \binom{n-t}{i}\left({c \over m}\right)^i\left(1-{c \over m}\right)^{n-t-i}\Bigr).
\end{align} 
Finally, we show that if at $j$-th iteration (having fixed $\Gamma(u_1) = C_1, \ldots, \Gamma(u_{j-1}) = C_{j-1}$ at the beginning) $\Gamma(u_j) = C_j$ is chosen so as to minimize $\expc[Y+Z\vert \Gamma(u_1)=C_1, \ldots, \Gamma(u_{j}) = C], C \in \binom{\mathcal{R}}{c}$, then in the final graph, which is no longer random, conditions (a) and (b) are met. To this end, we first observe that
\begin{align}
\nonumber
&\expc[Y+Z\vert \Gamma(u_1)=C_1, \ldots, \Gamma(u_{t-1}) = C_{t-1}]\\
\nonumber
 = &{\sum\limits_{C \in \binom{\mathcal{R}}{c}}\expc[Y+Z\vert \Gamma(u_1)=C_1, \ldots, \Gamma(u_{t}) = C] \over \binom{m}{c}}\\
&\geq \min \limits_{C \in \binom{\mathcal{R}}{c}}\expc[Y+Z\vert \Gamma(u_1)=C_1, \ldots, \Gamma(u_{t}) = C].
\end{align}
Hence, it follows that
\begin{align}
\label{exineq1}
\nonumber
&\min \limits_{C_1, \ldots , C_{n} \in \binom{\mathcal{R}}{c}}\expc[Y+Z\vert \Gamma(u_1)=C_1, \ldots, \Gamma(u_{n}) = C_{n}]\\
\nonumber
\leq &\min \limits_{C_1, \ldots , C_{n-1} \in \binom{\mathcal{R}}{c}}\expc[Y+Z\vert \Gamma(u_1)=C_1, \ldots, \Gamma(u_{n-1}) =C_{n-1}]\\
\nonumber
&\vdots\\
&\leq\min \limits_{C_1 \in \binom{\mathcal{R}}{c}}\expc[Y+Z\vert \Gamma(u_1)=C_1]
\leq \expc[Y+Z]
\leq {3 \over 4}.
\end{align}
Since $Y$ and $Z$ are integer valued random variables, the above essentially means that at the end, when all the neighbourhoods $\Gamma(u_1), \ldots, \Gamma(u_n)$ are fixed, $Y=0$ and $Z=0$. So, both the conditions (a) and (b) are met. Now, we have the following straight-forward algorithm to construct the bipartite graph.\\
\begin{algorithm}[H]
 \SetAlgoLined 
 \KwIn{Positive constants $c, k$, and sufficiently large $m$.}
 \KwOut{A bipartite graph $(\mathcal{L}, \mathcal{R}, \mathcal{E})$, where $\mathcal{L} = \{u_1, u_2, \ldots, u_n\}(n =  {m^{c-1+{1 \over k}}\over 4k^{c+1}})$ and $\mathcal{R} = \{v_1, v_2, \ldots, v_m\}$ such that $\Gamma(u_j) = C_j \in \binom{\mathcal{R}}{c}, 1\leq j\leq n$ meeting conditions (a) and (b).}
$\alpha ={nc \over m} -\sqrt{{n \over 2}\ln (4m)}$, and $\beta ={nc \over m} +\sqrt{{n \over 2}\ln (4m)}$;\\

 \For{$j \leftarrow 1$ \KwTo $n$}{
	$U_{j-1}= \{u_1, u_2,\ldots, u_{j-1}\}, min \leftarrow 1$\\
	\For{$C \in \binom{\mathcal{R}}{c}$}{
 
\begin{align*}
Y' \leftarrow \sum_{\substack{c+1\leq i\leq k}}~\sum_{\substack{u_j \in S \subset \mathcal{L}, \\ \lvert S\rvert =i}}~\sum_{\substack{S' \subset \mathcal{R}, \\ \lvert S'\rvert =i-1\\ \Gamma(U_{j-1}\cap S)\cup C \subseteq S' }}\left({\binom{i-1}{c}\over \binom{m}{c}}\right)^{i-\lvert U_{j-1}\cap S\rvert-1}
\end{align*}
\begin{align*}
Z & \leftarrow \sum\limits_{v \in \mathcal{R}}\Bigl(\sum\limits_{i=0}^{\alpha-\lvert U_{j-1}\cap \Gamma(v)\rvert-\lvert \{v\}\cap C \rvert-1} \binom{n-j}{i}\left({c \over m}\right)^i\left(1-{c \over m}\right)^{n-j-i}\\
&+\sum\limits_{i=\beta-\lvert U_{j-1}\cap \Gamma(v)\rvert-\lvert \{v\}\cap C \rvert+1}^{n-j} \binom{n-j}{i}\left({c \over m}\right)^i\left(1-{c \over m}\right)^{n-j-i}\Bigr)
\end{align*}
\If{$min > Y'+Z$}{ $\Gamma(u_j) = C$ \\
$min \leftarrow Y'+Z$ }
 
  }
 }
 \caption{}
\label{alg1}
\end{algorithm}
\noindent{\bf{\em Proof of correctness of the algorithm.}} At the beginning of $j$-th iteration, $\Gamma(u_1) = C_1, \ldots, \Gamma(u_{j-1}) = C_{j-1}$ are fixed and the algorithm selects $C = C_j$ which minimizes $Y'+Z$ for given $\Gamma(u_1) = C_1, \ldots, \Gamma(u_{j-1}) = C_{j-1}, \Gamma(u_j) = C$. Note that in (\ref{speccondvio}), $\expc[Y_{S, S'}\vert \Gamma(u_1)=C_1, \ldots, \Gamma(u_t) = C_t]$ is independent of particular choice of $C_i$ if $u_i \notin S$. So, in $j$-th iteration of the algorithm, while computing $Y'$, only those summands $\expc[Y_{S, S'}\vert \Gamma(u_1)=C_1, \ldots, \Gamma(u_j) = C]$ are considered for which $u_j \in S$. Hence, $Y' \leq \expc[Y_{S, S'}\vert \Gamma(u_1)=C_1, \ldots, \Gamma(u_j) = C]$, and particular choice of $C = C_j$ which minimizes $Y'+Z$ for given $\Gamma(u_1) = C_1, \ldots, \Gamma(u_{j-1}) = C_{j-1}, \Gamma(u_j) = C$ also minimizes $E[Y+Z\vert \Gamma(u_1) = C_1, \ldots, \Gamma(u_{j-1}) = C_{j-1}, \Gamma(u_j) = C]$; by (\ref{exineq1}), this also justifies setting $min$ to $1$ at the beginning of $j$-th iteration for $1 \leq j \leq n$. Hence  the proof follows from the discussion preceeding Algorithm \ref{alg1}.\par
\noindent{\bf{\em Runtime of the algorithm.}} Now, we present a coarse analysis of the algorithm which is sufficient to indicate that the algorithm runs in time $poly(m)$. For that, we first estimate the time required by the algorithm to compute $Y$\footnote{We consider RAM model (see \cite{MR95}), so addition, multiplication, and division are atomic operations}. Note that the time reqired to compute $\binom{m}{c}$ and $\binom{i-1}{c}$ is $O(m^2)$ (through dynamic programming); the exponentiation takes time $O(\log k)$, and these operations are done $O(kn^{k-1}m^{k-1})$ times to get the summation. So, the time required by the algorithm to compute $Y$ is $O(m^{(c+1)(k-1)+2})$. Similarly, in the case of computing $Z$, the binomial coefficients $\binom{n-i}{j}$ takes time $O(n^2)$ to be evaluated, exponentiations take time $O(\log n)$. So, the overall time requirement in this case is $O(mn^3\log n) = O(m^{3c-1} \log m)$.
The above two steps are done $O(nm^c)= O(m^{2c})$ times. So, the overall complexity of the algorithm is $O(m^{(k+1)(c+1)})$.
\section{Concluding Remarks}
\noindent{\bf{\em Limitations of the construction.}} It can be observed that the algorithm crucially depends on the fact that $k$ is a constant, and this limits its applicability to wider setting where $k$ is allowed to vary. Apart from being globally explicit, which, as discussed in the beginning, is a weaker notion of explicitness, the construction is on the slower side (even in terms of the number of edges which is $O(m^c)$), as indicated by the above analysis. One of the possible approaches to speed-up the algorithm is discussed below. 

\noindent{\bf{\em Towards derandomization in NC.}} It can be observed from the first part of Theorem \ref{cbcbound1} that the construction is in {\em RNC}, i.e., the construction can be carried out on a probabilistic Parallel Random Access Machine (PRAM) (see \cite{MR95} for definition) with $poly(m)$ many processors in constant time\footnote{In fact, it can be seen that the construction is in {\em ZNC} with the expected number of iterations at most $4$}. It is naturally interesting to investigate {\em NC}-derandomization of problems in {\em RNC} i.e., whether the same problem can be solved using a deterministic PRAM under same set of restrictions on resources. Two of the most commonly used techniques employed for such derandomization are the method of conditional expectation and the {\em method of small sample spaces} (see \cite{AS}); sometimes they are used together {\cite{BR91, MNN94}}.\par
While the method of conditional expectation performs a binary search (or more commonly a $d$-ary search) on the sample space for a good point, 
method of small sample spaces takes advantage of small independence requirement of random variables involved in the algorithm and constructs a small sized (polynomial in the number of variables) sample space which ensures such independence, and searches the sample space for a good sample point. Since the sample space is polynomial sized the search can be done in polynomial time, and hence leads to polynomial time construction.\par
 In the proof of Theorem \ref{cbcbound1} we used independence twice; in (\ref{indep1}) we used $k$-wise independence and in (\ref{chreq}) we used Hoeffding's inequality (Theorem \ref{hoef}) which requires the involved random variables $X_1, \ldots, X_n$ to be $n$-wise independent\footnote{Here we again point out that though the events and random variables are different in two cases, $k$-wise independence in choices of $\Gamma(u), u \in \mathcal{L}$ induces $k$-wise independence among random variables $X^{u}_{v}, u \in \mathcal{L}$ for fixed $v \in \mathcal{R}$}. However, such independence comes at the cost of a large sample space. More precisely, in \cite{ABI86} it was shown that in order to ensure $k$-independence among $n$ random variables the sample space size have to be $\Omega(n^{k \over 2})$. So, in case of Theorem \ref{cbcbound1} requirement on the size of sample space is huge ($\Omega(m^m)$). However, we want to point out here that the requirement of $n$-wise independence in Theorem \ref{cbcbound1} can be brought down to $O(\ln(m))$-wise independence with the help of following limited independence Cheronoff bound from \cite{BR94}. First we state the bound.
\begin{theorem}(\cite{BR94})
\label{cherlim}
Let $t\geq 4$ be an even integer. Suppose $X_1, \ldots, X_n$ be $t$-wise independent random variables taking values in $[0, 1]$. Let $X = X_1+ \cdots+X_n$, and $a > 0$. Then
$$
\pr\{\lvert X - \expc[X]\rvert \geq a\} \leq C_t \left({nt \over a^2}\right)^{t \over 2}
$$ , where $C_t$ is a constant depending on $t$, and $C_t < 1$ for $t \geq 6$.
\end{theorem}
Now, in inequality \ref{chreq}, we need $1 \over 2m $ in the r.h.s. This can be achieved by setting $t = 2\ln(2m)$ in the above theorem (for simplicity we assume $2\ln(2m)$ to be even), and $a = \sqrt{2en\ln(2m)}$. Hence, $O(\ln(m))$-wise independence in choosing $\Gamma(u)$ for $u \in \mathcal{L}$ is sufficient for the randomized construction (with somewhat inferior bound on the deviation of the degrees from the average).\par
In \cite{BR91, MNN94}, the authors developed frameworks for {\em NC}-derandomization of certain algorithms (notably, the set discrepancy problem of Spencer \cite{Spe94}) that require $O(\log^c(n))$-wise independence. One of the vital points of these frameworks is parallel computation of relevant conditional expectations for limited independence random variables in logarithmic time. In case of Algorithm \ref{alg1}, this means computation of $Y'$ and $Z$ by $poly(m)$ processors in $polylog(m)$ time under $O(\ln(m))$-wise independence among random choices of $\Gamma(u)$ for $u \in \mathcal{L}$. At present it is not clear to us how this can be achieved in the frameworks of \cite{BR91, MNN94} and seems to require more specialized technique.
\nopagebreak
\label{Bibliography}
\bibliographystyle{hep}  
\bibliography{derandomized_cbc}  

\end{document}